\documentclass[prd,aps,nofootinbib,superscriptaddress]{revtex4}
\usepackage{epsfig}
\usepackage{amsmath, slashed, mathtools}
\usepackage{xcolor}
\usepackage{appendix}
\usepackage{dsfont}
\usepackage[normalem]{ulem}

\usepackage{bm}
\usepackage{amsfonts}

\definecolor{pine}{rgb}{0.3, 0.5, 0.3}

\allowdisplaybreaks
\begin{document}

\title{Gravitational form factors in the perturbative limit }

\author{Qin-Tao Song}
\email[]{songqintao@zzu.edu.cn}
\affiliation{School of Physics, Zhengzhou University, Zhengzhou, Henan 450001, China}

\author{O. V. Teryaev}
\email[]{teryaev@jinr.ru}
\affiliation{Joint Institute for Nuclear Research, Joliot-Curie street 6, Dubna 141980, Russia}

\author{Shinsuke Yoshida}
\email[]{shinyoshida85@gmail.com}
\affiliation{Key Laboratory of Atomic and Subatomic Structure and Quantum Control (MOE),
Guangdong Basic Research Center of Excellence for Structure and Fundamental Interactions of Matter,
Institute of Quantum Matter, South China Normal University, Guangzhou 510006, China}
\affiliation{Guangdong-Hong Kong Joint Laboratory of Quantum Matter,
Guangdong Provincial Key Laboratory of Nuclear Science, Southern Nuclear Science Computing Center,
South China Normal University, Guangzhou 510006, China}

\date{\today}

\begin{abstract}
{Generalized distribution amplitudes (GDAs) have attracted significant attention in recent years due to their connection with the energy-momentum tensor (EMT) form factors (FFs). The GDAs can be experimentally
accessed through the study of  amplitudes in
$\gamma^{\ast} \gamma \to M_1 M_2$ and $\gamma^{\ast}  \to M_1 M_2 \gamma$, where $M_1M_2$ is a pseudoscalar meson pair such as $\pi \eta $ and $\eta \eta^{\prime}$. In this work, we calculate these amplitudes in the perturbative limit and express the extracted $M_1M_2$ GDAs in terms of meson distribution amplitudes that have been constrained by the previous experiments. Our explicit calculation verifies the existence of a new EMT FF  that
violates the conservation law of EMT when the hadronic matrix element of the EMT operator is
considered separately for each quark flavor. In addition, our result shows that the $M_1M_2$ GDAs are identical in $\gamma^{\ast} \gamma \to M_1 M_2$ and $\gamma^{\ast}  \to M_1 M_2 \gamma$, which confirms the universality of GDAs in the perturbative limit. In the future, the GDAs and the EMT FFs studied in this paper
can be probed at Belle II. Our study enhances the accessibility to the $P$-wave GDAs in $\gamma^{\ast} \gamma \to M_1 M_2$ and $\gamma^{\ast} \to M_1 M_2 \gamma$, and provides a promising approach for searching for exotic hybrid mesons in future experiments.}
\end{abstract}

\maketitle

\date{}

\section{Introduction}
\label{introduction}

Hadronic matrix elements of the energy-momentum tensor (EMT) are commonly
parameterized in terms of EMT form factors (FFs), which are also known as gravitational FFs.
The EMT FFs have attracted considerable interest because they
provide insight into the proton spin puzzle~\cite{Ji:1996ek}
and mechanical properties of hadrons~\cite{Polyakov:2018zvc,Burkert:2018bqq,Burkert:2023wzr,Lorce:2018egm,Dutrieux:2021nlz,Kumericki:2019ddg,Shanahan:2018nnv,Freese:2021czn,Li:2023izn,GarciaMartin-Caro:2023toa,Nair:2024fit,Lorce:2025ayr,Broniowski:2025ctl,Goharipour:2025lep}.
They also describe the interaction of hadrons with classical gravity and the manifestation of the equivalence principle which can be considered for quarks and gluons separately~\cite{Teryaev:1999su,Teryaev:2016edw}.

Generalized parton distributions (GPDs) and generalized distribution amplitudes(GDAs) serve as indirect probes of
the EMT FFs in the spacelike and timelike regions, respectively~\cite{Diehl:2003ny,Belitsky:2005qn, Boffi:2007yc, Goeke:2001tz,Polyakov:1998ze}.
The GDAs have been extensively studied
in hadron pair production processes such as
$\gamma^{\ast} \gamma  \to  h \bar{h}  $~\cite{Diehl:1998dk,Diehl:2000uv,Lorce:2022tiq} and  $\gamma^{\ast} \to h \bar{h} \gamma$~\cite{Lu:2006ut,BaBar:2015onb, Pire:2023kng, Pire:2023ztb}, where a perturbative treatment is valid
at large photon virtuality $Q^2$ and small invariant mass
squared $s$ of the meson pair.
These analyses can be extended to the $M_1M_2$ GDAs accessed
in the production of two different pseudoscalar mesons such as $\pi \eta $ and $\eta \eta^{\prime}$.
In contrast to the  $\pi \pi$ case,
there could be the $P$-wave component in the $M_1 M_2$ GDAs, which interestingly leads to the existence of a new EMT FF which
may be called the shear viscosity term $\Theta_3$\,\cite{Teryaev:2022pke}.
The $\Theta_3$ term has not been commonly considered
because it breaks the conservation law of hadronic matrix elements of EMT.
However, in principle, it could exist for a single quark flavor or gluon
as long as it vanishes when we take the sum for all the flavors and gluon.
At current stage, there is no clear evidence for the existence of the $\Theta_3$ term.
The smallness of its value may indicate a connection~\cite{Teryaev:2022pke} between the approximate validity of the equivalence principle for quarks and gluon individually and the low viscosity observed in the holographic framework~\cite{Kovtun:2004de}, with the latter further supported by a recent calculation involving black hole gravity~\cite{Lapygin:2025zhn}.

The $P$-wave GDAs are related to the study of exotic hybrid mesons. If a resonance is observed from the $P$-wave $M_1 M_2$ in
$\gamma^{\ast} \gamma \to M_1 M_2$ \cite{Anikin:2006du} and $\gamma^{\ast}  \to M_1 M_2 \gamma$ \cite{Pire:2023kng}, its quantum  number
$J^{PC}=1^{-+}$ cannot be described by the quark model.
Recently, several candidates
with $J^{PC}=1^{-+}$
have been reported by experiment
such as $\eta_1(1855)$~\cite{BESIII:2022riz,BESIII:2022iwi}, $\pi_1(1400)$\cite{IHEP-Brussels:1988iqi,E862:2006cfp},
$\pi_1(1600)$\cite{E852:1998mbq,COMPASS:2018uzl,JPAC:2018zyd,COMPASS:2021ogp} and
 $\pi_1(2015)$\cite{E852:2004gpn}.  However, further investigation is still needed for the confirmation of
 the $\pi_1$ states~ \cite{ParticleDataGroup:2022pth,Chen:2022asf}.
 In future, one can investigate these exotic states in the production of $M_1 M_2$, which is possible at Belle II and BESIII.
Furthermore, the measurements of these reactions can be used to study the hadronic light-by-light contribution to the muon's anomalous magnetic moment~\cite{Proceedings:2014ttc,Danilkin:2018qfn,Danilkin:2019mhd,Moussallam:2021dpk}.

Although the $\pi \pi$ GDA~\cite{Kumano:2017lhr} has been extracted from $\gamma^{\ast}  \gamma  \to \pi^0 \pi^0$~\cite{Belle:2015oin}, there are
no experimental measurements for the $M_1 M_2$ GDAs.
In light of the current situation, we would like to focus on an interesting way to indirectly
evaluate GDAs using meson distribution amplitudes(DAs).
It is known that the $\pi \pi$ GDA can be expressed in terms of pion DAs
in the amplitude for $\gamma^{\ast}  \gamma  \to \pi^+ \pi^-$  in the
kinematic region
$Q^2 \gg s \gg \Lambda_{\text{QCD}}^2$, which is referred to
as the perturbative limit~\cite{Diehl:1999ek}.
In this work, we apply the same technique to the helicity amplitudes for
$\gamma^{\ast} \gamma \to M_1 M_2$ and $\gamma^{\ast}  \to M_1 M_2 \gamma$
in order to
establish a relation between  $M_1 M_2$  GDAs and meson DAs in the perturbative limit.
The obtained relation will be used to
confirm the presence of a nonzero $\Theta_3$ term and, in addition,
to verify the universality of the GDAs
in $\gamma^{\ast} \gamma \to M_1 M_2$
and $\gamma^{\ast}  \to M_1 M_2 \gamma$.

The remainder of this paper is organized as follows:
In Section II, we provide a brief introduction to the $\pi\eta$ GDAs and define the kinematical
variables in $\gamma^{\ast}  \gamma  \to \pi \eta$.
In Section III, we present a detailed calculation of the helicity amplitude for $\gamma^{\ast}  \gamma  \to \pi \eta$.
The relation between GDAs and DAs is obtained in the perturbative limit.
In Section IV, we extend this analysis to
$\gamma^{\ast}  \to M_1 M_2 \gamma$.
The universality of the GDAs is discussed.
In Section V, we evaluate the EMT FFs in terms of the meson DAs. Section VI concludes the paper with a summary.

\section{$\pi \eta$ GDAs of $\gamma^{\ast}  \gamma  \to \pi \eta$  }
\label{kin}
We define the following kinematic variables for
the process of $\gamma^{\ast} (q) \gamma (q_1) \to \pi^0 (p)\eta(p_1)$,
\begin{align}
P=p+p_1, \hspace{4mm}
\Delta=p_1-p, \hspace{4mm}
\xi=\frac{p \cdot q_1 }{P \cdot q_1 }, \hspace{4mm}
q^2=-Q^2,  \hspace{4mm}
(q_1)^2=0, \hspace{4mm}
s=P^2, \hspace{4mm}
t=(q-p)^2, \hspace{4mm}
u=(q-p_1)^2.
\label{eqn:kinva}
\end{align}
We work in a frame where the virtual photon has
only nonzero $z$-component, $q=(0,0,0,Q)$.
For convenience,
we introduce
two lightcone vectors $n^{\mu}=(1,0,0,-1)/\sqrt{2}$ and $\bar{n}=(1,0,0,1)/\sqrt{2}$, and they are expressed in terms of  $q$ and $q_1$,
\begin{align}
 n=\frac{\sqrt{2}Q}{Q^2+s}q_1, \, \bar{n}=\frac{\sqrt{2}}{Q}q+\frac{\sqrt{2}Q}{Q^2+s}q_1.
\label{eqn:2lcs}
\end{align}
The light-cone components of a Lorentz vector
$a^{\mu}$ are defined as $a^+=a \cdot n$ and $a^-=a \cdot \bar{n}$.

\begin{figure}[htp]
\centering
\includegraphics[width=0.3\textwidth]{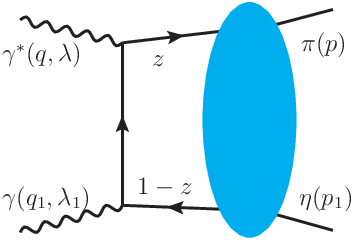}
\caption{The leading-twist $\pi\eta$ GDA is accessed in $\gamma^{\ast} \gamma \to \pi \eta $.}
\label{fig:etagda}
\end{figure}

If the QCD factorization ($Q^2 \gg s, \Lambda_{\text{QCD}}^2$) holds,  the amplitude for $\gamma^{\ast}  \gamma  \to \pi \eta$  can be
factorized into two parts, the subprocess $\gamma^{\ast} \gamma \to q \bar{q}$ as the hard scattering part and
the $\pi\eta$ GDA $\Phi^q_{\pi \eta}(z,\zeta, s)$ as the nonperturbative soft part.
The factorized formula is illustrated in Fig.\,\ref{fig:etagda}.
The $\pi\eta$ GDA is defined by~\cite{Anikin:2006du,Anikin:2004ja,Anikin:2004vc,Anikin:2005ke},
\begin{align}
\Phi^q_{\pi \eta}(z,\xi, s)
=\int \frac{dx^-}{2\pi} \,e^{-iz P^+ x^-}\,   \langle \eta(p_1) \pi(p)  | \,\bar{q}(x^-) \gamma^+ q(0)\, | 0 \rangle,
\label{eqn:gda}
\end{align}
where the Wilson line reduces to unity in
the lightcone gauge $A^+=0$
and $z$ indicates the momentum fraction carried by the quark
hadronizing into $\pi\eta$ pair.
The dependence on the factorization scale $\mu_F^2$, which is conventionally set as $\mu_F^2=Q^2$, is omitted in Eq.\,(\ref{eqn:gda})
for simplicity. The charge conjugation leads to
\begin{align}
\Phi^q_{\pi \eta}(z,\xi, s)=-\Phi^q_{\pi \eta}(\bar{z},\xi, s),
\label{eqn:gda-sys}
\end{align}
with $\bar{z} \equiv 1-z$. Unlike the $\pi^0 \pi^0$ GDA, the $\pi \eta$ GDA does not satisfy
$\Phi^q(z,\xi, s)=\Phi^q(z,\bar{\xi}, s)$
followed by the interchange of identical particles.
In the asymptotic limit $Q^2 \to \infty$, the $\pi \eta$  GDA takes the form
\begin{align}
\Phi^q_{\pi \eta}(z,\xi, s)= 10z\bar{z}C_1^{(3/2)}(2z-1) \sum_{l=0}^2 B_{1l}(s)P_l(\cos \theta),
\label{eqn:gda-as}
\end{align}
where $l$ represents the orbital angular momentum of the meson pair, and
$\theta$ is the polar angle of the meson pair in the $\gamma^{\ast} \gamma$ center of mass frame which can be expressed in terms of the parameter $\xi$,
\begin{align}
\beta \cos \theta =2\xi-1-\frac{m_{\pi}^2-m_{\eta}^2}{s},  \,  \beta=\frac{\lambda^{\frac{1}{2}}(s,m_\pi^2,m_\eta^2)}{s},
\label{eqn:theta-po}
\end{align}
where $\lambda(s,m_\pi^2,m_\eta^2)$ is the  Kallen function.

At the lowest order with respect to $\alpha_s$,
the leading-twist amplitude for $\gamma^{\ast}  \gamma  \to \pi \eta$ survives only in the case that the incoming photons have the same helicity,
and it is expressed as~\cite{Anikin:2006du},
\begin{align}
M_{\lambda \lambda_1}=\frac{e^2 }{2} \delta_{\lambda \lambda_1}  \sum_q e_q^2  \int_0^1 dz \frac{2z-1}{z\bar{z}} \Phi^q_{\pi \eta}(z,\xi, s),
\label{eqn:amp-gda}
\end{align}
where the helicities of the virtual and real photons are denoted as
$\lambda$ and $\lambda_1$, respectively.

\section{$\pi\eta$  GDAs in perturbative limit }
\label{GDAs}

\begin{figure}[htp]
\centering
\includegraphics[width=0.5\textwidth]{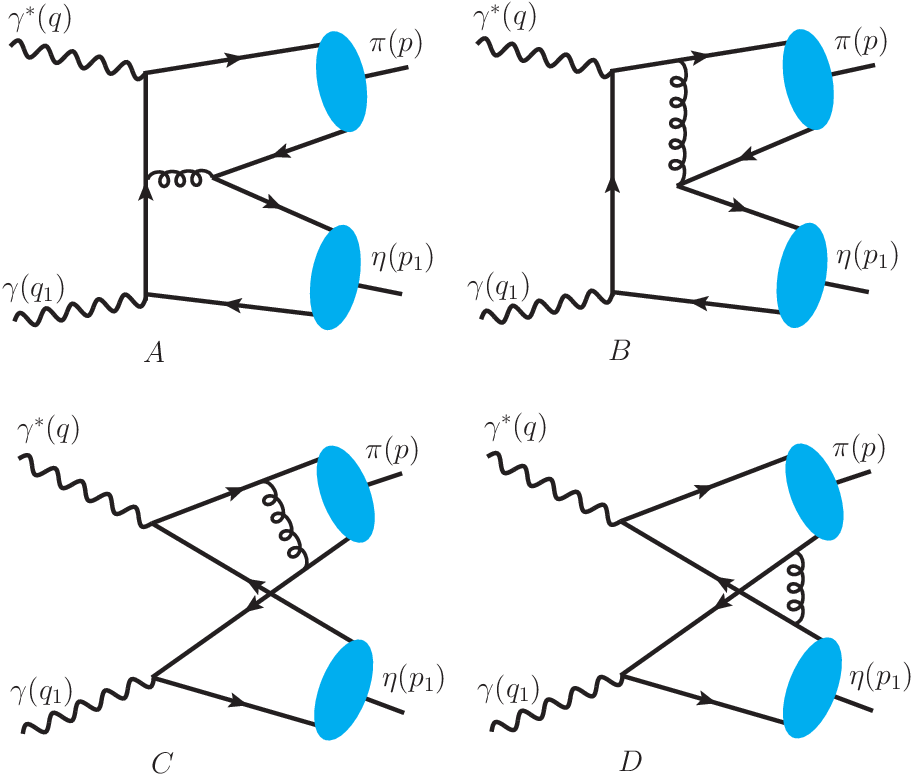}
\caption{Feynman diagrams for $T_{\lambda \lambda_1}^{1q}$ in $\gamma^{\ast}  \gamma  \to \pi^0 \eta$.}
\label{fig:num0}
\end{figure}

In the perturbative limit $Q^2 \gg s \gg \Lambda_{\text{QCD}}^2$, the amplitude for $\gamma^{\ast}  \gamma  \to \pi^+ \pi^-$
can be expressed in terms of the hard scattering amplitude
and soft pion DAs~\cite{Brodsky:1981rp, Diehl:1999ek}.
This approach can be extended to
the process $\gamma^{\ast}  \gamma  \to \pi \eta $,
and the helicity amplitude is given by
\begin{align}
M_{\lambda \lambda_1}=  \sum_q   \int_0^1 dx dy \left[ T_{\lambda \lambda_1}^{1q}(x,y,s,t,Q^2)  \phi_{\eta}^q(y)
+ T_{\lambda \lambda_1}^{2q}(x,y,s,t,Q^2)  \phi_{\eta}^g(y) \right]\phi_{\pi}^q(x) ,
\label{eqn:amp-pieta}
\end{align}
where $T_{\lambda \lambda_1}^{1q}$  is the parton level amplitude
for $\gamma^{\ast}  \gamma  \to q \bar{q}+ q \bar{q}$,
and the sum
runs over the quark flavors $u$ and $d$
due to isospin conservation.
The function $\phi_M^q(z)$ denotes
the DA for a neutral pseudoscalar meson $M$, such as $\pi^0$ and $\eta^{(\prime)}$, and is defined as
\begin{align}
\phi_M^q(z) = i \int \frac{dx^-}{2\pi} \,e^{-iz p^+ x^-}\,   \langle  M(p)  | \,\bar{q}(x^-) \gamma^+ \gamma_5 q(0)\, | 0 \rangle,
\label{eqn:da}
\end{align}
where $z$ is the momentum fraction carried by the quark
hadronizing into $M$.
$\phi_M^q(z)$ can be written
in terms of its eigenfunctions, Gegenbauer  polynomials,
\begin{equation}
\begin{aligned}
\phi_M^{u,d}(z)&= 6f_M^{u,d}z\bar{z}
\sum_{i=0} a_{2i}^M C_{2i}^{3/2}(2z-1),\\
\phi_{\eta^{(\prime)}}^s(z)&= 6f_{\eta^{(\prime)}}^s z\bar{z}\sum_{i=0} \tilde{a}_{2i}^{\eta^{(\prime)}} C_{2i}^{3/2}(2z-1),
\label{eqn:da-ge}
\end{aligned}
\end{equation}
where $a_{0}^M=\tilde{a}_{0}^{\eta^{(\prime)}}=1$, and $f_M^q$ is the $M$-meson decay constant
for the quark flavor $q$. The isospin symmetry
implies $f_{\pi}=\sqrt{2} f^u_{\pi}=-\sqrt{2} f^d_{\pi} $
for $\pi^0$. The $s$-quark DA exists for $\eta^{(\prime)}$ meson.
The Gegenbauer coefficients are different between $u/d$ and $s$,
and are denoted by $a_{2i}^{\eta^{(\prime)}}$
and $\tilde{a}_{2i}^{\eta^{(\prime)}}$, respectively.

In addition, we introduce an alternative definition of the  $\eta^{(\prime)}$ DAs,
\begin{align}
\phi_{\eta^{(\prime)}}^{1,8}(z) = i \int \frac{dx^-}{2\pi} \,e^{-iz p^+ x^-}\,   \langle  \eta^{(\prime)}(p)  | J_{5}^{1,8\,+} (x^-;0)| 0 \rangle,
\label{eqn:da2}
\end{align}
where the $SU(3)$ flavor-singlet current $J_{5}^{1\mu}$
and the flavor-octet current $J_{5}^{8\mu}$
are given by
\begin{equation}
\begin{aligned}
J_{5}^{1\mu}(x^-;0) &=\frac{1}{\sqrt{3}} \left[ \bar{u}(x^-) \gamma^{\mu} \gamma_5 u(0) + \bar{d} (x^-)\gamma^{\mu} \gamma_5 d (0)+ \bar{s}(x^-) \gamma^{\mu} \gamma_5 s(0) \right ],  \\
J_{5}^{8\mu}(x^-;0)  &=\frac{1}{\sqrt{6}} \left[ \bar{u}(x^-) \gamma^{\mu} \gamma_5 u(0) + \bar{d} (x^-)\gamma^{\mu} \gamma_5 d (0)-2 \bar{s}(x^-) \gamma^{\mu} \gamma_5 s(0) \right ].
\label{eqn:cur18}
\end{aligned}
\end{equation}
Similarly to Eq.~\eqref{eqn:da-ge},
these DAs can be expanded in terms of Gegenbauer  polynomials,
\begin{equation}
\begin{aligned}
\phi_{\eta^{(\prime)}}^1(z)= 6f_{\eta^{(\prime)}}^1 z\bar{z}\sum_{i=0} \bar{a}_{2i} C_{2i}^{3/2}(2z-1), \\
\phi_{\eta^{(\prime)}}^8(z)= 6f_{\eta^{(\prime)}}^8 z\bar{z}\sum_{i=0} \hat{a}_{2i} C_{2i}^{3/2}(2z-1).
\label{eqn:da-18}
\end{aligned}
\end{equation}
Following the convention in~\cite{Kroll:2002nt, Kroll:2012gsh}, we assume that the Gegenbauer coefficients $\bar{a}_{2i}$ and $\hat{a}_{2i}$ are the same for the $\eta$ and $\eta'$ mesons, respectively.
The decay constants $f_{\eta^{(\prime)}}^{1,8}$
and the Gegenbauer coefficients $\bar{a}_{2i}$
and $\hat{a}_{2i}$ are different from those defined for
each quark flavor in~\eqref{eqn:da-ge}. $f_{\eta}^1$ and $f_{\eta^{\prime}}^8$ describe the deviation of the $\eta$ and $\eta^{\prime}$
mesons from their naive quark content, respectively, and they are responsible for the mixing of $\eta$ and $\eta^{\prime}$ mesons (see e.g.\cite{Klopot:2012hd}).

From Eqs.~\eqref{eqn:da},~\eqref{eqn:da2}
and ~\eqref{eqn:cur18}, one can show the following relations,
\begin{equation}
\begin{aligned}
\phi_{\eta^{(\prime)}}^u(z)&=\phi_{\eta^{(\prime)}}^d(z)=\frac{1}{\sqrt{6}} \left[ \sqrt{2} \phi_{\eta^{(\prime)}}^1(z)+\phi_{\eta^{(\prime)}}^8(z) \right], \\
\phi_{\eta^{(\prime)}}^s(z)&= \frac{1}{\sqrt{3}} \left[  \phi_{\eta^{(\prime)}}^1(z)-\sqrt{2} \phi_{\eta^{(\prime)}}^8(z) \right].
\label{eqn:da-ge-18uds}
\end{aligned}
\end{equation}
Taking the first moments of these equations with respect to $z$,
we obtain the relations among the
decay constants~\cite{Feldmann:1998sh,Charng:2006zj,Ball:2007hb,Agaev:2014wna,Bali:2021qem},
\begin{equation}
\begin{aligned}
f_{\eta^{(\prime)}}^u=&f_{\eta^{(\prime)}}^d=\frac{1}{\sqrt{6}} \left[  \sqrt{2}f_{\eta^{(\prime)}}^{1} +f_{\eta^{(\prime)}}^{8}\right ],  \\
f_{\eta^{(\prime)}}^s=&\frac{1}{\sqrt{3}} \left[ f_{\eta^{(\prime)}}^{1} -\sqrt{2}f_{\eta^{(\prime)}}^{8}\right ].
\label{eqn:decay-eta}
\end{aligned}
\end{equation}
As a result, the relations among the Gegenbauer coefficients
are given by
\begin{equation}
\begin{aligned}
a^{\eta^{(\prime)}}_{2i}&=\frac{\sqrt{2} f_{\eta^{(\prime)}}^1
\bar{a}_{2i} +f_{\eta^{(\prime)}}^8 \hat{a}_{2i} }{\sqrt{2} f_{\eta^{(\prime)}}^1  +f_{\eta^{(\prime)}}^8  } , \\
\tilde{a}^{\eta^{(\prime)}}_{2i}&
=\frac{ f_{\eta^{(\prime)}}^1 \bar{a}_{2i}
-\sqrt{2} f_{\eta^{(\prime)}}^8 \hat{a}_{2i} }{
f_{\eta^{(\prime)}}^1  - \sqrt{2}f_{\eta^{(\prime)}}^8  }.
\label{eqn:da-ge-sin}
\end{aligned}
\end{equation}
We will evaluate the shear viscosity term $\Theta_3$
in terms of $a^{\eta^{(\prime)}}_{n}$ and
$\tilde{a}^{\eta^{(\prime)}}_{n}$
because we are interested in the existence of a nonzero
$\Theta_3$ for each quark flavor and its cancellation
among all the relevant quark flavors.
However, the expression in terms of $\bar{a}_n$ and
$\hat{a}_n$ is also important when we take into account
the scale evolution with respect to $\mu_F$ as we will discuss
below, and Eq.~\eqref{eqn:da-ge-sin} enables us to
switch these two expressions.

In Eq.~\eqref{eqn:amp-pieta}, $T_{\lambda \lambda_1}^{2q}$ denotes the amplitude of $\gamma^{\ast}  \gamma  \to q \bar{q}+ gg$.
The hadronization of the gluon pair into $\eta^{(\prime)}$
is described by the gluon DA,
which is defined as~\cite{Charng:2006zj, Ball:2007hb,Agaev:2014wna,Kroll:2002nt,Ali:2003kg}
\begin{align}
 p_1^+ \int \frac{dx^-}{2\pi} \,e^{-iz p_1^+ x^-}\,   \langle  \eta^{(\prime)}(p_1)  | \,A_a^{[ \alpha}(x^-)  A_b^{\beta]}(0)   \, | 0 \rangle
 =\frac{-1}{3\sqrt{3}}    \epsilon_T^{\alpha \beta} \frac{\delta_{ab}}{8}  \frac{\Psi_M^g(z)}{z \bar{z}},
\label{eqn:da-gl0}
\end{align}
where the symbol $[\mu\nu]$ denotes antisymmetrization of a tensor,
$t^{[\mu \nu]}=\tfrac{1}{2}(t^{\mu \nu}-t^{\nu \mu})$,
and $\epsilon_T^{\alpha \beta}$ is
given by
\begin{align}
\epsilon_T^{\alpha \beta} =\frac{\epsilon^{\alpha \beta \mu \nu }p_{1\mu} q_{1\nu} }{p_1\cdot q_1}.
\label{eqn:anti}
\end{align}
The gluon DA is also expressed in terms of Gegenbauer polynomials,
\begin{align}
\Psi_{\eta^{(\prime)}}^g(z)=f_{\eta^{(\prime)}}^1z^2(1-z)^2\sum_{i=1} b_{2i} C_{2i-1}^{5/2}(2z-1),
\label{eqn:da-gl}
\end{align}
where we assume that the Gegenbauer coefficients
are the same for $\eta$ and $\eta^{\prime}$~\cite{Kroll:2002nt, Kroll:2012gsh}.
 The gluon DA of $\eta^{(\prime)}$ vanishes
in the asymptotic limit $Q^2\to \infty$.
For convenience, we adopt the convention $\phi_{\eta^{(\prime)}}^g(z) \equiv \Psi_{\eta^{(\prime)}}^g(z)/(z\bar{z})$ .
Charge conjugation leads to $\Psi_{\eta^{(\prime)}}^g(z)=  -\Psi_{\eta^{(\prime)}}^g(\bar z)$ and
\begin{align}
\int_0^1 dz\,\Psi_{\eta^{(\prime)}}^g(z)=\int_0^1 dz\,\phi_{\eta^{(\prime)}}^g(z)=0.
\label{eqn:da-gl1}
\end{align}

The meson DAs depend on the factorization scale $\mu_F$, whose
evolution is governed by the ERBL
equation~\cite{Lepage:1979zb,Efremov:1979qk}.
The evolution equations for the Gegenbauer coefficients
$a^{\pi}_{2i}$ and $\hat{a}_{2i}$  are given by the same form,
\begin{align}
F_{n}(\mu_F)= F_{n}(\mu_0)L^{\gamma_n^{q q}/ \beta_0},
\hspace{5mm}F_n=\{a^{\pi}_{n},\hat{a}_{n}\},
\label{pion-evo}
\end{align}
where $L=\alpha_s(\mu_0)/ \alpha_s(\mu_F)$, $\beta_0=11-2N_f/3$, and $\gamma_n^{q q}$ is the  anomalous dimension,
\begin{align}
\gamma_n^{q q}=C_F\left[3+\frac{2}{(n+1)(n+2)}-4 \sum_{i=1}^{n+1} \frac{1}{k}\right].
\label{qq-ad}
\end{align}
The Gegenbauer coefficients $\bar{a}_n$ and $b_{n}$
mix under the evolution~\cite{Baier:1981pm, Kroll:2002nt},
\begin{equation}
\begin{aligned}
& \bar{a}_n\left(\mu_F\right)=a_{n}^{+}\left(\mu_0\right)L^{\gamma_n^{+} / \beta_0}+\rho_n^{-} a_{n}^{-}\left(\mu_0\right)L^{\gamma_n^{-} / \beta_0}, \\
& b_{n}\left(\mu_F\right)=\rho_n^{+} a_{n}^{+}\left(\mu_0\right)L^{\gamma_n^{+} / \beta_0}+a_{n}^{-}\left(\mu_0\right)L^{\gamma_n^{-} / \beta_0},
\label{evol18}
\end{aligned}
\end{equation}
where
\begin{equation}
\begin{aligned}
\gamma_n^{ \pm}=&\frac{1}{2}\left[\gamma_n^{q q}+\gamma_n^{g g} \pm \sqrt{\left(\gamma_n^{q q}-\gamma_n^{g g}\right)^2+4 \gamma_n^{q g} \gamma_n^{g q}}\right], \\
\rho_n^{+}=&6 \frac{\gamma_n^{g q}}{\gamma_n^{+}-\gamma_n^{g g}}, \quad \rho_n^{-}=\frac{1}{6} \frac{\gamma_n^{q g}}{\gamma_n^{-}-\gamma_n^{q q}} .
\end{aligned}
\end{equation}
The anomalous dimensions are given by
\begin{equation}
\begin{aligned}
\gamma_n^{q g}=&C_F \frac{n(n+3)}{3(n+1)(n+2)}, \\
\gamma_n^{g q}=&N_f \frac{12}{(n+1)(n+2)}, \\
\gamma_n^{g g}=&\beta_0+N_c\left[\frac{8}{(n+1)(n+2)}-4 \sum_{k=1}^{n+1} \frac{1}{k}\right].
\end{aligned}
\end{equation}
The initial conditions of $a_{n}^{\pm}\left(\mu_0\right)$
are simply written in terms of $\bar{a}_n\left(\mu_0\right)$ and
$b_{n}\left(\mu_0\right)$ by setting $\mu_F=\mu_0$ in Eq.~\eqref{evol18}.
Note that we do not take into account the scale dependence
of the singlet decay constant $f_{\eta^{(\prime)}}^1$,
because it is the order of $\alpha_s^2$ and ignored in this study.

To express the twist-2 $\pi \eta$ GDAs in terms of meson DAs, we need to calculate the helicity amplitudes   of Eq.~\eqref{eqn:amp-pieta}, where the helicities of the incoming photons are identical due to Eq.~(\ref{eqn:amp-gda}).
There are four types of Feynman diagrams contributing to
$T_{\lambda \lambda_1}^{1q}$ as shown in Fig.\,\ref{fig:num0}, and the
additional diagrams can be found by
particle interchange from these four diagrams~\cite{Diehl:1999ek}.
We choose the light-cone gauge $A^+=0$ for the calculation
in which the gluon propagator takes the form
\begin{align}
\frac{i \delta_{a b}}{l^2+i \epsilon}\left(-g^{\mu \nu}+\frac{l^\mu q_1^{ \nu}+q_1^{ \mu} l^\nu}{l \cdot q_1}\right),
\end{align}
where $l$ is the momentum of the gluon. If we expand the hard amplitudes  in powers of $s/Q^2$, the  diagrams $A$ and $C$,
the diagram $B$, and the diagram $D$
are order of $1/Q^2$, $1/s$, and $s/Q^4$, respectively.
Thus we only keep the contribution from $B$
in the perturbative limit of
$Q^2 \gg s \gg \Lambda_{\text{QCD}}^2$~\cite{Diehl:1999ek}.
The first term in Eq.~\eqref{eqn:amp-pieta} is then evaluated as
\begin{align}
e^2\delta_{\lambda \lambda_{1}}
\sum_qe_q^2\Bigl(-{8\pi \alpha_s\over 9}\Bigr)
\int_0^1  \frac{dx dy }{\bar{x}ys}
\Big\{  \frac{u+(2y-1)t}{\bar{y}(u+yt)}  \frac{(2-x)u+yt}{\bar{x}u+yt}  +\frac{t-(2x-1)u}{x(t+\bar{x}u)}  \frac{(1+y)t+\bar{x}u}{yt+\bar{x}u}
\Big\} \phi_{\pi}^q(x) \phi_{\eta}^q(y),
\label{eqn:amp-das1}
\end{align}
In Eq.~\eqref{eqn:amp-das1}, the first (second) part originates
from the diagrams where the gluon propagator
connects the quark-antiquark pair hadronizing into $\pi$ ($\eta$) meson. Neglecting the terms of order $\mathcal{O} (s/Q^2)$, we have $t=-(1-\xi)Q^2$ and $u=-\xi Q^2$.
Thus, Eq.~\eqref{eqn:amp-das1} can be re-expressed as
\begin{align}
e^2\delta_{\lambda \lambda_{1}}
\sum_qe_q^2\Bigl(-{8\pi \alpha_s\over 9}\Bigr)
\int_0^1 dz     \frac{2z-1}{z\bar{z}}    \int_0^1 \frac{dx }{s} & \left[    \theta(z-\xi)  \frac{\bar{\xi}}{z-\xi}  \frac{z+\bar{x}\xi}{z-x \xi} \frac{\phi_{\pi}^q(x) }{\bar{x}}
\phi_{\eta}^q(\tfrac{\bar{z}}{\bar{\xi}}) \right.  \nonumber \\
  & \left. +\theta(\xi-z)  \frac{\xi}{z-\xi}   \frac{\bar{z}+ \bar{x} \bar{\xi}}{\bar{z}-x \bar{\xi}} \frac{\phi_{\eta}^q(x) }{\bar{x}}
 \phi_{\pi}^q(\tfrac{z}{\xi}) \right ],
\label{eqn:amp-b1b2}
\end{align}
where $\theta(x)$ is the step function, and the symmetry $\phi_M^q(x)=\phi_M^q(\bar{x})$ is taken into account.
Comparing  Eq.\,(\ref{eqn:amp-b1b2}) with Eq.\,(\ref{eqn:amp-gda}),
the contribution from the first term in Eq.~\eqref{eqn:amp-pieta} to
the quark GDAs is extracted,
\begin{equation}
\begin{aligned}
\hat{\Phi}^{q}_{\pi \eta}(z,\xi, s)\Bigr|_{\rm quark\ DAs}
=-{16\pi \alpha_s\over 9}
& \Bigg\{    \theta(z-\xi)  \frac{\bar{\xi}}{z-\xi}  \int_0^1 \frac{dx }{s} \frac{z+\bar{x}\xi}{z-x \xi} \frac{\phi_{\pi}^q(x) }{\bar{x}}
         \phi_{\eta}^q(\tfrac{\bar{z}}{\bar{\xi}}) \\
          &+\theta(\xi-z)  \frac{\xi}{z-\xi}  \int_0^1 \frac{dx }{s} \frac{\bar{z}+ \bar{x} \bar{\xi}}{\bar{z}-x \bar{\xi}} \frac{\phi_{\eta}^q(x) }{\bar{x}} \phi_{\pi}^q(\tfrac{z}{\xi})
 \Bigg\}.
\label{eqn:gdadu}
\end{aligned}
\end{equation}

\begin{figure}[ht]
\centering
\includegraphics[width=0.5\textwidth]{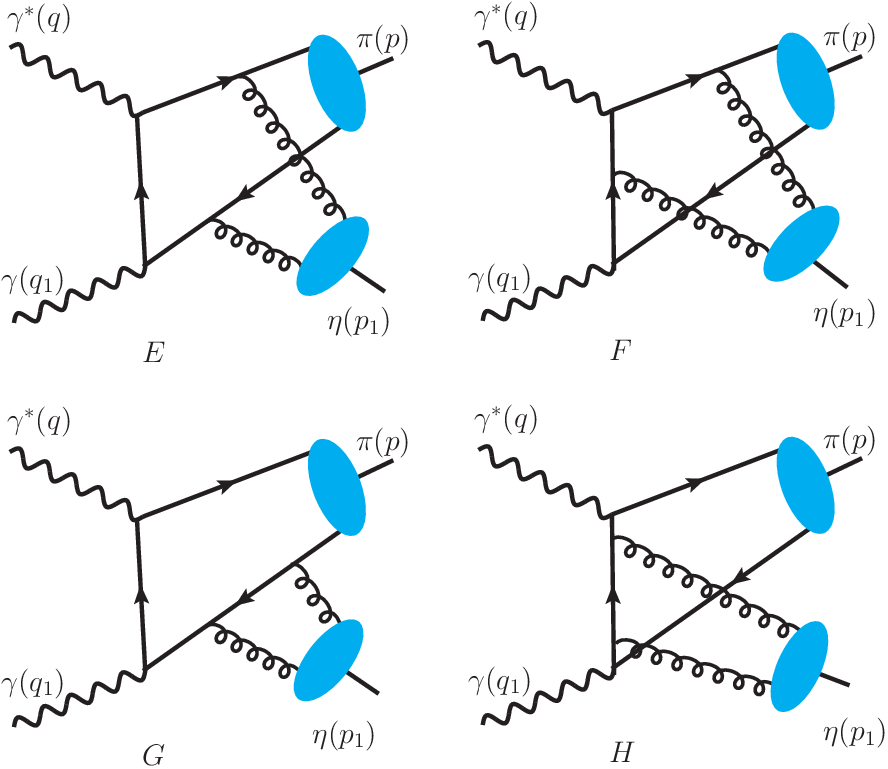}
\caption{Feynman diagrams for $T_{\lambda \lambda_1}^{2q}$ in $\gamma^{\ast}  \gamma  \to \pi \eta$}
\label{fig:gluon}
\end{figure}

We next consider $T_{\lambda \lambda_1}^{2q}$ in
Eq.~\eqref{eqn:amp-pieta}
associated with the $\eta$ gluon DA.
There are also four types of Feynman diagrams contributing to
$T_{\lambda \lambda_1}^{2q}$ as shown in Fig.~\ref{fig:gluon}.
Only Feynman diagrams of $E$ and $G$ contribute to the  amplitude at the order of $1/s$, however, the $F$ and $H$  diagrams are suppressed by $s/Q^2$,  and are therefore neglected in the perturbative limit. The second term of Eq.~\eqref{eqn:amp-pieta}  is then evaluated as
\begin{equation}
\begin{aligned}
 e^2\delta_{\lambda \lambda_{1}}
\sum_qe_q^2\Bigl(-{2\pi \alpha_s\over 9\sqrt{3}}\Bigr)
\int_0^1 \frac{dx dy }{\bar{x}x\bar{y}ys} &\Bigg\{  \frac{(2y-1)t+ (2x-1)u  }{\bar{y}t+\bar{x}u} \frac{ \bar{y} y t+ \left [x- (2x-1) y \right]u   }{yt+xu }   \\
&+\frac{\bar{y}^2 \left[t-(2x-1) u\right]}{t+\bar{x}u}
 + \frac{\bar{y}^2\left[t+(2x-1) u\right]}{t+xu}
\Bigg\} \phi_{\pi}^q(x) \phi_{\eta}^g(y).
\label{eqn:amp-gl}
\end{aligned}
\end{equation}
The first part in Eq.~\eqref{eqn:amp-gl} originates from the diagram $E$. The second (third) part originates from the diagram $G$
in which the gluon pair couples to the antiquark (quark)
hadronizing into $\pi$, and
they give identical contributions due to the symmetry $\phi_{\pi}^q(x)=\phi_{\pi}^q(\bar{x})$.
We then extract the corresponding GDAs  from the amplitude of
Eq.~\eqref{eqn:amp-gl},
\begin{equation}
\begin{aligned}
\hat{\Phi}^{q}_{\pi \eta}(z,\xi, s)\Bigr|_{\rm quark-gluon\ DAs}
={4\pi \alpha_s\over 9\sqrt{3}}
\frac{\xi}{s} &\Bigg\{        \int_{S_1} \frac{dy }{\bar{y}y} \frac{y^2-(2y-1) z +
\xi \bar{y}y}{(z-y-\xi \bar{y})(z- \bar{\xi} y)}\phi_{\eta}^g(y) \phi_{\pi}^q(x)\\
          & -\int_0^1 dy  \frac{ \bar{y} }{y} \left [ \frac{\theta(\xi-z) }{z-\xi}\phi_{\pi}^q(\tfrac{z}{\xi}) -  \frac{\theta(\xi-\bar{z}) }{\bar{z}-\xi}\phi_{\pi}^q(\tfrac{\bar{z}}{\xi})  \right ]\phi_{\eta}^g(y)
\Bigg \},
\label{eqn:gda-gl}
\end{aligned}
\end{equation}
where the condition Eq.~\eqref{eqn:da-gl1} is used
to simplify the expression. The three parts in Eq.~\eqref{eqn:gda-gl} correspond directly to those in Eq.~\eqref{eqn:amp-gl}.
In Eq.~\eqref{eqn:gda-gl}, we need to regard $x$  as a function of $y$ and $z$,  $x=(z-y\bar{\xi})/\xi $, and $S_1$ represents the integration boundary of $y$, defined by the condition $|x+y-1| + |x-y| \leq 1$.

The $\pi\eta$ GDA is just the sum of Eqs.~\eqref{eqn:gdadu} and \eqref{eqn:gda-gl},
\begin{align}
\hat{\Phi}^{q}_{\pi \eta}(z,\xi, s)=
\hat{\Phi}^{q}_{\pi \eta}(z,\xi, s)\Bigr|_{\rm quark\ DAs}
+\hat{\Phi}^{q}_{\pi \eta}(z,\xi, s)\Bigr|_{\rm quark-gluon\ DAs}.
\label{eqn:gdato}
\end{align}
Note that $\hat{\Phi}^{q}_{\pi \eta}(z,\xi, s)$ is not exactly same as
$\Phi^{q}_{\pi \eta}(z,\xi, s)$ because it contains the
components that violate the charge conjugation symmetry
given in Eq.~\eqref{eqn:gda-sys}.
We can re-express $\hat{\Phi}^{q}_{\pi \eta}(z,\xi, s)$ as
\begin{align}
\hat{\Phi}^{q}_{\pi \eta}(z,\xi, s)=\frac{1}{2} \left[ \hat{\Phi}^{q}_{\pi \eta}(z,\xi, s)- \hat{\Phi}^{q}_{\pi \eta}(\bar{z},\xi, s)  \right]
+\frac{1}{2} \left[ \hat{\Phi}^{q}_{\pi \eta}(z,\xi, s)+ \hat{\Phi}^{q}_{\pi \eta}(\bar{z},\xi, s)  \right],
\label{eqn:sys}
\end{align}
where only the first part satisfies the symmetry of Eq.\,(\ref{eqn:gda-sys}), and the second part does not contribute to the amplitude of Eq.~\eqref{eqn:amp-gda}
due to the existence of the prefactor $(z-\bar{z})/(z \bar{z})$.
Therefore, in the perturbative limit the $\pi\eta$ GDA is given by
\begin{align}
\Phi^{q}_{\pi \eta}(z,\xi, s)=\frac{1}{2} \left[ \hat{\Phi}^{q}_{\pi \eta}(z,\xi, s)- \hat{\Phi}^{q}_{\pi \eta}(\bar{z},\xi, s)  \right].
\label{eqn:sys1}
\end{align}
The formula (\ref{eqn:sys1}) could give logarithmic singularities around $z=\xi$ and $z=1-\xi$ for a specific form of the DAs
like the asymptotic form $\phi^q_{\pi}(x)\sim x(1-x)$\cite{Diehl:1999ek},
which means that the relation between GDAs and DAs is not well-defined in the whole region of $z$.
Fortunately, the singularities are integrable and canceled in the moment of the GDAs as we will see in the section V.
We regard this as a fact that the moment of the GDAs is insensitive to the soft contribution which spoils
the factorized formula (\ref{eqn:sys1}) in the perturbative limit.
More careful treatment was discussed in \cite{Diehl:1999ek} by introducing a cutoff to regularize the singularities.
It was shown that the cutoff dependence turns to a power correction $O(\Lambda_{QCD}^2/s)$ after taking the moment
and, therefore, the moment of the GDAs which is expressed in terms of DAs is trustworthy as long as the power correction is negligible.

\section{Universality of GDAs}

The $\pi\eta$  GDAs can be probed by
a spacelike photon in  $\gamma^{\ast}  \gamma  \to \pi \eta$.
Currently, there are no experimental facilities capable of
testing the $\pi\eta$ GDAs  of Eq.\,(\ref{eqn:sys1})
in the perturbative limit. However, the $\pi\eta$  GDAs can
also be accessed in $\gamma^{\ast}   \to \pi \eta \gamma$, which can be investigated at Belle II in the perturbative limit. The virtual photon is timelike in this process and
the helicity amplitudes are also given by Eq.\,(\ref{eqn:amp-gda})\,\cite{Lu:2006ut,Pire:2023kng}.
Therefore, the universality of GDAs can be tested
by comparing the results from
$\gamma^{\ast} \to \pi \eta \gamma$ and $\gamma^{\ast}  \gamma  \to \pi \eta$, analogous to the universality of hadron GPDs in timelike and spacelike hard exclusive processes~\cite{Mueller:2012sma}.

We define the following variables to describe the amplitude for $\gamma^{\ast} (q)  \to \pi^0 (p)\eta(p_1)\gamma (q_1) $,
\begin{align}
q^2=Q^2, \hspace{5mm}
(q_1)^2=0,  \hspace{5mm}
\xi=\frac{p \cdot q_1 }{(p+p_1) \cdot q_1 },  \hspace{5mm}
(p+p_1)^2=s,    \hspace{5mm}
(q_1+p_1)=\hat{t},  \hspace{5mm}
(q_1+p)=\hat{u},
\label{eqn:kinti}
\end{align}
where  $\hat{t}$ and $\hat{u}$ are the squared invariant masses for the final meson-photon pairs, and they are positive unlike the variables $t$ and $u$ in Eq.~\eqref{eqn:kinva}.
The process $\gamma^{\ast} \to \pi \eta \gamma$ is related to
$\gamma^{\ast}  \gamma  \to \pi \eta$ through the crossing symmetry.
Our explicit calcualtion shows that the amplitudes for
$\gamma^{\ast} \to \pi \eta \gamma$ can also be described by
Eqs.~\eqref{eqn:amp-das1} and \eqref{eqn:amp-gl} after the following replacements,
\begin{align}
t \to \hat{t}, \, u \to \hat{u}.
\end{align}
$\hat{t}$ and $\hat{u}$ are respectively given by $\hat{t}=(1-\xi)Q^2$
and $\hat{u}=\xi Q^2$.
The $\pi\eta$ GDAs extracted from the helicity amplitudes
of the timelike process $\gamma^{\ast}  \to \pi \eta \gamma$
are consistent with Eqs.~\eqref{eqn:gdadu}, \eqref{eqn:gda-gl} and \eqref{eqn:sys1} which are derived from the spacelike process $\gamma^{\ast}  \gamma  \to \pi \eta$ in the perturbative limit.
Thus, our calculation verifies the universality of GDAs in this limit.

At Belle II, the kinematic variables can reach
$Q^2\sim100$ GeV$^2$ and  $s\sim10$ GeV$^2$ in $\gamma^{\ast}   \to \pi \eta \gamma$, and  the perturbative limit $Q^2 \gg s \gg \Lambda_{\text{QCD}}^2$ is sufficiently satisfied. Therefore,
our formulas for $\pi\eta$ GDAs
can be tested experimentally in the near future.

\section{Gravitational form factors}
\label{shear}

The EMT for a single quark flavor $T^{\mu \nu}_q$  is defined as
\begin{align}
    T^{\mu\nu}_{q}(0)=\frac{i}{2}
    \overline{q}(0)\gamma^{\{\mu}
    \overset{\leftrightarrow}{D}\!\!\!\!\!\phantom{D}^{\nu\}} q(0),
\label{EMTdf}
\end{align}
where $t^{\{\mu \nu\}}=\tfrac{1}{2}(t^{\mu \nu}+t^{\nu \mu})$. The first moment of the $\pi\eta$ GDA  corresponds to the timelike matrix element of EMT operator~\cite{Diehl:1998dk,Diehl:2000uv,Polyakov:1998ze},
\begin{align}\label{eqn:sum-m}
\int^1_0 dz \rho_z \Phi^q_{\pi\eta}(z, \xi, s)
=\frac{2}{(P^+)^2}
\left \langle \eta(p_1) \pi(p) \left| T^{++}_q(0) \right|0\right \rangle,
\end{align}
which is expressed in terms of the transition EMT FFs \cite{Pagels:1966zza,Teryaev:2022pke},
\begin{align}
\left \langle \eta(p_1) \pi(p) \left| T^{\mu \nu}_q(0) \right|0\right \rangle
=\frac{1}{2} \left[ \Theta_1^q(s)(sg^{\mu \nu}-P^{\mu}P^{\nu})+    \Theta_2^q(s)  \Delta^{\mu} \Delta^{\nu}
+  \Theta_3^q(s) P^{\{\mu} \Delta^{\nu  \} }\right].
\label{eqn:emta}
\end{align}
The $\Theta_3^q$ term does not appear in the $\pi \pi$ case
and the positive $\Theta_1^q(s=0)$ corresponds to the
stability condition of pion \cite{Goeke:2007fp,Hudson:2017xug}.
It is also notable that $\Theta_1^q(s=0)=\Theta_2^q(s=0)$ has been
obtained for $\pi \pi$ by the model calculations~\cite{Donoghue:1991qv,Polyakov:1999gs,Polyakov:1998ze}.
The new term $\Theta_3^q(s)$ is associated with the $P$-wave component of
GDA, which violates
the conservation law $\left \langle \eta(p_1) \pi(p) \left| T^{\mu \nu}_q(0) \right|0\right \rangle P_{\mu}=0$~\cite{Teryaev:2022pke}.
However, this violation vanishes when summing  $\Theta_3$ over
all quark flavors and the gluon.
Now these EMT FFs can be expressed in terms of the meson DAs
using the extracted $\pi\eta$ GDAs in Eq.~\eqref{eqn:sys1},
\begin{equation}
\begin{aligned}
\Theta_1^q  &=-\frac{c}{s}  \int dx dy \left[\frac{1+\bar{x}+y  }{\bar{x}y} \phi_{\eta}^q(y) +  \frac{\tilde{c}y }{\bar{x}x} \phi_{\eta}^g(y)  \right] \phi_{\pi}^q(x),  \\
\Theta_2^q  &=-\frac{c}{s}\int dx dy  \left[ \frac{ 1+x+\bar{y}}{\bar{x}y} \phi_{\eta}^q(y) -  \frac{\tilde{c}y }{\bar{x}x} \phi_{\eta}^g(y)   \right ]\phi_{\pi}^q(x),  \\
\Theta_3^q  &=\frac{2c}{s}\int dx dy \left[  \frac{x-\bar{y}}{\bar{x}y} \phi_{\eta}^q(y)  +  \frac{\tilde{c}y }{\bar{x}x} \phi_{\eta}^g(y)\right]\phi_{\pi}^q(x),
\label{eqn:thes}
\end{aligned}
\end{equation}
where $c= -8\pi \alpha_s/9$ and $\tilde{c}= 1/(4\sqrt{3})$.
Substituting Eqs.~\eqref{eqn:da-ge} and \eqref{eqn:da-gl} into
Eq.~\eqref{eqn:thes}, the timelike EMT FFs can be written as
\begin{equation}
\begin{aligned}
\Theta_1^q  =&-\frac{c f^q_\pi  }{2s} \Big\{  6\left [  5 +4(a_2^{\pi}+a_2^{\eta} ) +3a_2^{\pi} a_2^{\eta} \right ]f^q_\eta   +\tilde{c}  (1+a_2^{\pi})b_2 f^1_{\eta}
\Big\},  \\
\Theta_2^q  =&-\frac{cf^q_\pi }{2s} \Big \{6 \left [  7 +8(a_2^{\pi}+a_2^{\eta} )+9a_2^{\pi} a_2^{\eta} \right ]f^q_\eta   - \tilde{c} (1+a_2^{\pi})b_2 f^1_{\eta} \Big \},  \\
\Theta_3^q =& \frac{cf^q_\pi}{s} \sum_{i=1 }  \Big[ 6(a_{2i}^{\pi}-a_{2i}^{\eta}) f^q_\eta +  \tilde{c}  (1+\sum_{j=1}a_{2j}^{\pi})  b_{2i} f^1_{\eta}  \Big],
\label{eqn:the2a}
\end{aligned}
\end{equation}
where $q=u$ or $d$ denotes the quark flavor, and the Gegenbauer  polynomials with $n\geq3$ are neglected for $\Theta_1^q$ and $\Theta_2^q$ due to lengthy expressions.
These formulas can simply be extend
to $\pi \eta^{\prime}$ by replacing the indices
$\eta$ with $\eta^{\prime}$.
If the asymptotic limit $Q^2\to \infty$ is taken, we obtain $7\Theta_1^q=5\Theta_2^q$ and find that the $\Theta_3^q$ term vanishes.
However, if the general expressions of meson DAs are adopted, the first term of $\Theta_3^q$
arises when the quark DA of the $\pi$ meson differs from that of the $\eta$ meson, and the second term will  be nonzero provided that the gluon DA does not vanish $(b_{2i}^{\eta} \neq 0)$, as illustrated by Eq.~\eqref{eqn:the2a}.

If we set $\mu_F=\sqrt{30}$ GeV,
recent studies predicts $a_2^{\pi}\left(\mu_F\right)\sim 0.16$ on average~\cite{Cloet:2013tta,Chernyak:1981zz, Ball:2006wn,Radyushkin:2009zg,Arthur:2010xf,Agaev:2010aq,Bakulev:2012nh,Mikhailov:2021znq,Kroll:2012sm,Raya:2015gva, Bali:2018spj, RQCD:2019osh,LatticeParton:2022zqc,Zhong:2021epq,Shi:2021nvg,Gao:2022vyh}, where the Eq.~\eqref{pion-evo} is used with the input $a_2^{\pi}\sim 0.25$ at $\mu_0=1$ GeV.
The authors of Ref.~\cite{Kroll:2012gsh} obtain $\hat{a}_2\left(\mu_0\right)=-0.05$,  $\bar{a}_2\left(\mu_0\right)=-0.12$, and $b_{2}\left(\mu_0\right)=19$
from a combined analysis of the experimental measurements  by the CLEO Collaboration \cite{CLEO:1997fho} and BABAR Collaboration~\cite{BaBar:2011nrp}. After incorporating the evolution effects described in Eqs.~\eqref{pion-evo} and \eqref{evol18}, we have $\hat{a}_2\left(\mu_F\right)=-0.032$,  $\bar{a}_2\left(\mu_F\right)=-0.027$, and $b_2\left(\mu_F\right)=7.6$ for the quark octet, singlet and gluon DAs, respectively.
We take the $\eta$ and $\eta^{\prime}$ decay constants from Ref.~\cite{Bali:2021qem},  and finally obtain the coefficients $a_2^{\eta}\left(\mu_F\right) \sim -0.03$ and $a_2^{\eta^{\prime}}\left(\mu_F\right) \sim -0.03$   using
Eq.~\eqref{eqn:da-ge-sin}. One can also infer from Eq.~\eqref{eqn:the2a}
that for the $\Theta_3^q$ term, the contribution of the $\eta^{\prime}$
gluon DA is much larger than that of the $\eta$ gluon DA due to $f^1_{\eta^{\prime}} \gg f^1_\eta$.
Thus, the existence of the $\Theta_3^q$  term seems quite plausible for
the $\pi \eta$ and $\pi \eta^{\prime}$ pairs.

Moreover, with some model assumptions \cite{Teryaev:2022pke}
one can express the famous ratio of viscosity to entropy density in terms
of $\Theta_3^q/\Theta_2^q$ which appears to be about $0.055$. It is slightly smaller than the bound \cite{Kovtun:2004de} equal to $1/4\pi \approx 0.08$, but
mentioned model assumptions cannot pretend for high accuracy and require further studies, especially in the timelike channel.
Using the isospin symmetry relations for meson DAs, we  find $\Theta_3^u(s)=-\Theta_3^d(s)$ for the $\pi \eta$ and $\pi \eta^{\prime}$  pairs.
Consequently, $\Theta_3$ term vanishes when summing over quark flavors, and the conserved hadronic matrix elements of EMT is recovered.

\begin{figure}[ht]
\centering
\includegraphics[width=0.7\textwidth]{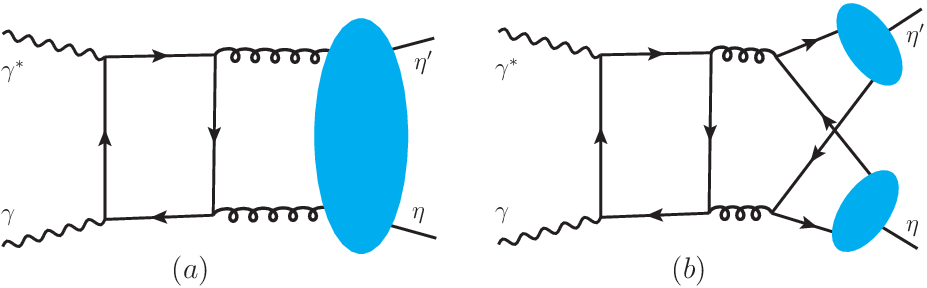}
\caption{(a) Gluon GDA is accessed in the higher order corrections to the amplitudes of $\gamma^{\ast} \gamma \to \eta^{\prime} \eta$~\cite{Kivel:1999sd}, and the factorization condition $Q^2 \gg s, \, \Lambda_{\text{QCD}}^2$ is satisfied; (b) Gluon GDA  is extracted from the higher order corrections to the amplitudes of $\gamma^{\ast} \gamma \to \eta^{\prime} \eta$ in the perturbative limit. }
\label{fig:hcs-gl}
\end{figure}

For the
$\eta^{\prime} \eta$ pair, Eq.~\eqref{eqn:thes} will be slightly modified,
\begin{equation}\label{eqn:2eta2}
\begin{aligned}
\Theta_1^q|_{\eta^{\prime} \eta}  &=-\frac{c}{s}  \int dx dy \frac{1+\bar{x}+y  }{\bar{x}y} \phi_{\eta}^q(y)\phi_{\eta^{\prime}}^q(x) -\frac{c\tilde{c} }{s}  \int dx dy \left[\frac{y }{\bar{x}x} \phi_{\eta^{\prime}}^q(x) \phi_{\eta}^g(y)   +   \frac{x }{\bar{y}y}\phi_{\eta^{\prime}}^g(x) \phi_{\eta}^q(y) \right],  \\
\Theta_2^q|_{\eta^{\prime} \eta}  &=-\frac{c}{s}\int dx dy  \frac{ 1+x+\bar{y}}{\bar{x}y} \phi_{\eta}^q(y) \phi_{\eta^{\prime}}^q(x)+\frac{c\tilde{c} }{s}\int dx dy  \left [ \frac{y }{\bar{x}x}\phi_{\eta^{\prime}}^q(x) \phi_{\eta}^g(y) +\frac{x }{\bar{y}y}\phi_{\eta^{\prime}}^g(x) \phi_{\eta}^q(y) \right],  \\
\Theta_3^q|_{\eta^{\prime} \eta}  &=\frac{2c}{s}\int dx dy  \frac{x-\bar{y}}{\bar{x}y} \phi_{\eta}^q(y) \phi_{\eta^{\prime}}^q(x)+\frac{2c\tilde{c}}{s}\int dx dy
\left[ \frac{y }{\bar{x}x} \phi_{\eta^{\prime}}^q(x) \phi_{\eta}^g(y)-\frac{x }{\bar{y}y}\phi_{\eta^{\prime}}^g(x) \phi_{\eta}^q(y) \right],
\end{aligned}
\end{equation}
where the quark flavor  can be $u$, $d$, or $s$.
We obtain $\Theta_3^u(s)=\Theta_3^d(s)$  using the isospin symmetry relation. The $\Theta_3$ term  should vanish when we sum over the quark flavors and the gluon,
 \begin{align}
\sum_{i=q,g} \Theta_3^i(s)=0.
\label{eqn:eta2}
\end{align}
The gluon GDA will appear in the amplitudes of
 $\gamma^{\ast} \gamma \to \eta^{\prime} \eta $~\cite{Kivel:1999sd} and $\gamma^{\ast}  \to \eta^{\prime} \eta \gamma$ when the higher-order corrections are included, and one of the typical Feynman  diagrams is depicted in Fig.~\ref{fig:hcs-gl}(a). In this work,
the gluonic contribution $\Theta_3^g$ is identically zero, and the existence of a nonzero $\Theta_3^g$ and its cancellation can be also investigated through the higher-order corrections to $\gamma^{\ast} \gamma \to \eta^{\prime} \eta $ and $\gamma^{\ast}  \to \eta^{\prime} \eta  \gamma$ in the perturbative limit, and we also show one of its typical Feynman diagrams in  Fig.~\ref{fig:hcs-gl}(b), which could be addressed in a future study.

In the future, the process $\gamma^{\ast} \to M_1 M_2  \gamma$ can be measured at Belle II in the perturbative limit,
which allows us to test the GDAs of Eq.~\eqref{eqn:sys1} and the EMT FFs of Eqs.~\eqref{eqn:thes} and \eqref{eqn:2eta2} experimentally.
At Belle II, the GDAs can be also accessed in  $\gamma^{\ast} \gamma \to M_1 M_2 $, although the kinematics in this case  do not satisfy the  perturbative limit. Given that the maximum values of $Q^2$ and $s$ in recent measurements~\cite{Belle:2015oin} are approximately 25 GeV$^2$ and 4 GeV$^2$, respectively, the kinematic boundary is close to satisfying the perturbative limit. Therefore, Eqs.~\eqref{eqn:sys1}, \eqref{eqn:thes} and \eqref{eqn:2eta2} can still serve as a boundary constraint in the extraction of GDAs.

The existence of  $\Theta_3^q(s)$  also indicates that there are $P$-wave GDAs for $\pi \eta$ and $\eta^{\prime} \eta $, enabling the search for exotic resonances through  the $P$-wave production of these meson pairs in $\gamma^{\ast} \gamma \to M_1 M_2$~\cite{Anikin:2006du,Anikin:2004ja} and $\gamma^{\ast}  \to M_1 M_2 \gamma$~\cite{Pire:2023kng}, which is possible at Belle II and BESIII. Recently, the  exotic  resonance $\eta_1(1855)$  was observed   by BESIII  through
 the $P$-wave analysis of  $ \eta^{\prime} \eta$ in  $J/\Psi \to \eta^{\prime} \eta  \gamma $ \cite{BESIII:2022riz,BESIII:2022iwi}.
Given the similarity between $J/\Psi $ and $\gamma^{\ast}$, it is promising to search for
$\eta_1(1855)$ in $\gamma^{\ast} \gamma \to  \eta^{\prime} \eta  $ and $\gamma^{\ast}  \to  \eta^{\prime} \eta \gamma$.

\section{Summary}
\label{sum}
In the perturbative limit  $Q^2 \gg s \gg \Lambda_{\text{QCD}}^2$, the helicity amplitudes of $\gamma^{\ast} \gamma \to M_1 M_2$ and $\gamma^{\ast}  \to M_1 M_2 \gamma$ can be factorized into
the hard scattering amplitudes and soft meson DAs even when $M_1 M_2$ is a pair of different mesons like $\pi \eta $ and $\eta^{\prime} \eta$.
This fact suggests a possible connection between the $M_1 M_2$ GDAs
and the meson DAs, particularly the gluon DA associated with
$\eta^{(\prime)}$ production.
We have derived the formulas for GDAs in terms of the quark and the gluon
DAs, and confirmed the universality of GDAs between
$\gamma^{\ast} \gamma \to M_1 M_2$ and $\gamma^{\ast}  \to M_1 M_2 \gamma$ in the perturbative limit.
The derived formulas allow us to express
the timelike transition EMT FFs in terms of the meson DAs whose
parameters have been constrained by the previous experiments.
We have verified the existence of a new EMT FF $\Theta_3^q$ which does not exist for the $\pi \pi$ case,
which confirms the anticipated appearance of exotic quantum numbers,
being the counterpart of naive T-violation in spacelike channel making, in turn, the contact with a dissipative nature of viscosity.
Although this new EMT FF violates
the conservation law of EMT
when its hadronic matrix element  is
considered for a single quark flavor,
our result ensures that the conservation law is restored for
$\pi \eta^{(\prime)} $ after
summing over all the relevant quark flavors.
At Belle II, the measurement of $\gamma^{\ast} \to M_1 M_2 \gamma$ satisfies the condition of the perturbative limit,
making it possible to test the $M_1 M_2$ GDAs and EMT FFs
obtained in this work experimentally.
 Furthermore, the obtained GDAs and EMT FFs can serve as boundary constraints for extracting $M_1 M_2$ GDAs from $\gamma^{\ast} \gamma \to M_1 M_2$ at Belle II and $\gamma^{\ast} \to M_1 M_2 \gamma$ at BESIII.
Since the $\Theta_3^q$ term originates from  the $P$-wave components of
the GDAs, our study suggests that it is feasible to search for exotic resonances through
the $P$-wave production of $M_1 M_2$ in $\gamma^{\ast} \gamma \to M_1 M_2$ and $\gamma^{\ast}  \to M_1 M_2 \gamma$.

\section*{Acknowledgments}
We acknowledge useful discussions with   Markus Diehl, Yang Li, C\'edric Lorc\'e, Sergey Mikhailov, Bernard Pire, Georgy Prokhorov and Valentin Zakharov. Qin-Tao Song was supported by the National Natural Science Foundation
of China under Grant Number 12005191. O.V. Teryaev is thankful to PiFi program of CAS for support
and Prof. Xurong Chen for hospitality at IMP (Huizhou).



\end{document}